\documentclass[acmsmall,screen,nonacm]{acmart}

\AtBeginDocument{}

\setcopyright{none}
\AtBeginDocument{\settopmatter{printacmref=true,printccs=true,printfolios=true}}
\acmYear{2026}
\acmDOI{}
\acmISBN{}

\makeatletter
\let\hyxmp@parse@acmart\relax
\makeatother
\hypersetup{pdfcontactemail={yilingl@illinois.edu},pdfcontactcountry={USA}}

\usepackage{booktabs}
\usepackage{colortbl}
\usepackage{enumitem}
\usepackage{float}
\usepackage{longtable}
\usepackage{microtype}
\usepackage{multirow}
\usepackage{needspace}
\usepackage{pifont}
\usepackage{placeins}
\usepackage{tabularx}
\usepackage{tabularray}
\usepackage{tcolorbox}
\usepackage{wrapfig}
\usepackage{xcolor}
\usepackage{xspace}
\newcommand{\distance}{5pt}
\definecolor{codexrow}{HTML}{EAF3FF}
\definecolor{claudecoderow}{HTML}{FFF1E6}
\definecolor{kimicoderow}{HTML}{ECF8EE}
\definecolor{orchestrationrow}{HTML}{DDECF2}
\definecolor{governancerow}{HTML}{E7EBDD}
\definecolor{contextrow}{HTML}{F0E2EE}
\definecolor{mergerow}{HTML}{F4E9CF}
\newcommand{\parabf}[1]{\noindent\textbf{#1}}

\newcommand{\researchq}[1]{\textbf{RQ#1}}
\newcommand{\taskdash}{\mbox{-}}
\newcolumntype{L}[1]{>{\raggedright\arraybackslash}p{#1}}
\newcolumntype{Y}{>{\raggedleft\arraybackslash}X}
\newtcolorbox{takeawaybox}{
  boxsep=0pt,
  before skip=6pt,
  after skip=6pt,
  colback=gray!8,
  colframe=gray!65,
  boxrule=0.5pt,
  arc=2mm,
  left=3pt,
  right=3pt,
  top=3pt,
  bottom=3pt
}

\title[When Sub-Agents Work in Parallel]{When Sub-Agents Work in Parallel: The Promises and Pitfalls of Dynamic Concurrency in Long-Horizon Coding Tasks}

\author{Han Li}
\affiliation{
  \institution{Nanjing University}
  \country{China}
}
\email{231220161@smail.nju.edu.cn}

\author{HanHaoNing Li}
\affiliation{
  \institution{Central University of Finance and Economics}
  \country{China}
}

\author{Ziqian Jiang}
\affiliation{
  \institution{Nanjing University}
  \country{China}
}

\author{Yiling Lou}
\affiliation{
  \institution{University of Illinois at Urbana-Champaign}
  \country{USA}
}
\email{yilingl@illinois.edu}

\begin{document}
\raggedbottom

\begin{abstract}
As coding agents advance toward increasingly complex and long-horizon software development tasks, dynamic concurrency has emerged as a promising mechanism for scaling agent execution. Unlike predefined multi-agent workflows, dynamic concurrency allows a main agent to decide during execution whether, when, and how to spawn and coordinate concurrent sub-agents. However, its effectiveness, efficiency, and failure characteristics remain poorly understood, particularly for long-horizon coding tasks. In this work, we present the first systematic study of native dynamic concurrency in three frontier coding agents: Codex, Claude Code, and Kimi Code. We conduct controlled comparisons with dynamic concurrency enabled or disabled across 354 tasks from four long-horizon benchmarks and one bounded benchmark, yielding 2,124 executions and costing over \$20K.  Our results show that dynamic concurrency benefits most on long-horizon tasks, while bounded tasks often see little or negative improvement. Meanwhile, concurrency substantially increases resource consumption, with token usage reaching 1.41–3.31 times that of sequential execution. Through trajectory analysis, we develop a taxonomy of concurrency failure patterns comprising four main categories, 13 subcategories, and 28 specific patterns. Finally, we identify conditions under which concurrency provides concrete benefits, including complementary parallel work, alternative solution exploration, and independent validation. Together, our findings reveal both the promise and limitations of dynamic concurrency and provide empirical guidance for enabling more effective dynamic collaboration among multiple sub-agents.

\end{abstract}


\maketitle

\FloatBarrier
\section{Introduction}
\label{sec:introduction}
The recent advances of Large Language Models (LLMs) have enabled a remarkable progress in LLM-based coding agents~\cite{yang2024sweagent,liu2026agentsurvey}, which demonstrate increasingly strong capabilities across a broad range of critical software engineering tasks, from code generation~\cite{chen2021codex} and debugging to repository-level issue resolution. As these agents are applied to increasingly complex and long-horizon tasks, however, a single agent often faces substantial limitations in its context, reasoning, and execution capacity. In particular, long-horizon software development may involve extensive code exploration, implementation, testing, and verification~\cite{bouzenia2025executionagent} over prolonged trajectories, motivating the need to scale agent execution beyond a single sequential trajectory~\cite{ding2026nl2repobench,yang2026programbench}.

Multi-agent systems provide a promising path toward such scaling. By decomposing a complex task into subtasks and delegating them to multiple sub-agents, these approaches can increase execution capacity, enable parallel exploration of alternative solutions, and reduce end-to-end latency. Recent studies and agent systems have demonstrated the potential of multi-agent collaboration and parallel execution across a variety of complex reasoning and software engineering tasks~\cite{qian2024chatdev,hong2023metagpt,wu2023autogen,kimi2026k25}, suggesting that concurrency can serve as an important mechanism for scaling coding agents to more demanding workloads.

In fact, the earlier coding agents (e.g., TRAE~\cite{trae2025agent} and  OpenHands~\cite{wang2024openhands}) often adopt a \textit{static concurrency} paradigm, which rely on predefined workflows and dedicated orchestration harnesses to coordinate the collaboration between sub-agents. For example, the number and roles of sub-agents, their task assignments, communication patterns, and synchronization points are often specified in advance for a particular workflow. While effective in well-structured settings, such a static concurrency requires developers to anticipate how a task should be decomposed and parallelized before execution. This rigidity makes it difficult to accommodate diverse and evolving tasks whose decomposition, dependencies, and opportunities for parallelism may become apparent only as execution unfolds, limiting the flexibility and generality of these approaches.

\begin{figure*}[t]
  \centering
  \includegraphics[width=\textwidth,pagebox=cropbox]{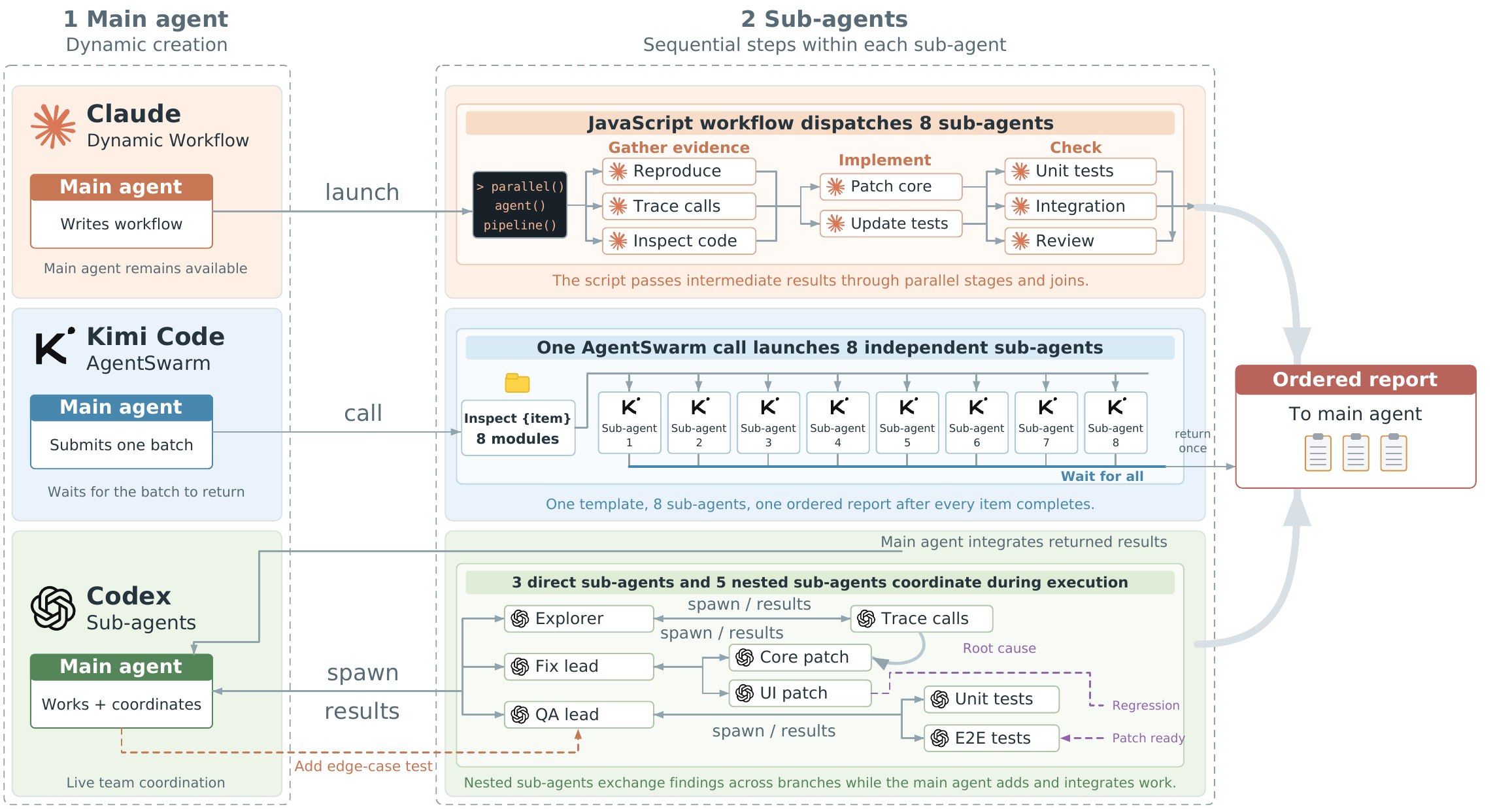}
  \caption{Overview of the dynamic concurrency mechanisms used by the three coding agents in our study.}
  \Description{Separate dashed panels show main agents on the left and sub-agents on the right across three tool rows. Claude executes a scripted sequence of parallel stages and joins. Kimi launches a batch of independent sub-agents and waits for one ordered report. Codex coordinates direct and nested sub-agents while the main agent continues working and integrates returned results.}
  \label{fig:study-overview}
\end{figure*}

A recent trend among frontier coding agents is moving beyond such predefined parallel workflows toward \textit{dynamic concurrency}. Rather than following a fixed orchestration, the main agent dynamically determines during execution whether, when, and how to spawn concurrent sub-agents based on the evolving task state. Figure~\ref{fig:study-overview} illustrates the dynamic concurrency in three frontier coding agents (e.g., Codex multiagent Version2~\cite{openaiCodexSubagents}, Claude Code Dynamic Workflow~\cite{anthropicDynamicWorkflows}, and Kimi Code AgentSwarm~\cite{moonshotKimiCode}).
With dynamic concurrency enabled, a main agent dynamically constructs an execution plan to orchestrate sub-agents while remaining active to collect their results, reconcile shared interfaces or files, and verify the integrated artifact. Sub-agents are instantiated and executed on demand according to instructions issued by the main agent. For example, in Claude Code, the main agent automatically generates a workflow script that schedules parallel stages for multiple sub-agents, specifies their explicit dependencies, and coordinates their joins. Similarly, in Kimi Code, the main agent launches a batch of sub-agent contexts instantiated from a shared template with task-specific assignments, waits for their completion, and returns an ordered report. In Codex, the main agent supports both direct and nested delegation of sub-agents while continuing its own execution and exchanging messages with sub-agents.

While extensive efforts have evaluated coding agents and analyzed their failure patterns, existing studies have largely focused on either single-agent systems~\cite{jimenez2024swebench,openai2024swebenchverified,liu2025automatedfailures,bouzenia2025repairagent,zhang2024autocoderover} or multi-agent systems organized around predefined collaboration protocols~\cite{qian2024chatdev,hong2023metagpt,huang2024agentcoder,islam2024mapcoder,chen2024coder,du2024multiagentdebate,wang2024mixtureofagents}. Moreover, existing evaluations predominantly target bounded coding tasks, such as SWE-bench~\cite{jimenez2024swebench,xia2025agentless}, where the expected outputs typically involve relatively small patches (e.g., only dozens of lines of code). In contrast, dynamic concurrency is particularly intended to support long-horizon and open-ended coding tasks, such as repository-level generation~\cite{zhang2026repozero,yang2026programbench,ding2026nl2repobench} and software release development~\cite{li2026loopsbench}, where the coding agents are expected to generate tens of thousands of lines of code per task. Therefore, it remains unclear how the emerging dynamic concurrency mechanisms in frontier coding agents affect their effectiveness and efficiency, particularly on long-horizon coding tasks.

To fill this knowledge gap, we perform the first study on the native dynamic concurrency mechanisms in frontier coding agents, including Codex multiagent Version2~\cite{openaiCodexSubagents}, Claude Code Dynamic Workflow~\cite{anthropicDynamicWorkflows}, and Kimi Code AgentSwarm~\cite{moonshotKimiCode}. In particular, we evaluate these coding agents with dynamic concurrency on a wide range of long-horizon coding tasks, including 226 repository generation tasks (e.g., RepoZero~\cite{zhang2026repozero}, ProgramBench~\cite{yang2026programbench}, NL2Repo~\cite{ding2026nl2repobench}) and 28 sustained repository evolution tasks (e.g., LoopsBench~\cite{li2026loopsbench}). These long-horizon benchmarks span mean reference implementation or gold-change sizes of 0.8--54.4 thousand lines across different difficulty groups, providing broad implementation scopes for evaluating dynamic concurrency. To further enable a comparison between long-horizon coding tasks and bounded coding tasks, we also include 100 bounded coding tasks from the classic SWE-bench Verified~\cite{openai2024swebenchverified}. In total, our evaluation yields 2,124 executions across a range of task complexities and execution horizons, with 11.2 billion recorded tokens corresponding to \$20,780.02 at standard API rates.

Built upon the evaluation results, we perform both quantitative and qualitative to analyze the promises and failure patterns of coding agents with dynamic concurrency in long horizon-coding tasks. In  (\researchq{1}), we first systematically compare the effectiveness and efficiency of coding agents with dynamic concurrency or without dynamic concurrency (i.e., sequential mode). Interestingly, we find that dynamic concurrency does not consistently improve end-to-end task success rates, e.g., the resolution effects range from an 11.9 percentage point decrease to a 2.3 point increase, while mean token use reaches 1.41 to 3.31 times compared to the sequential mode. In particular, dynamic concurrency delivers pass rate gains on long-horizon benchmarks, especially LoopsBench, while providing no such gains on the bounded tasks (SWE-bench Verified). In  (\researchq{2}), we further analyze the delegation and coordination behavior within each task progresses. In particular, we observe that dynamic concurrency is increasingly extended in more challenging tasks, e.g., the number of created sub-agents increases along with the task difficulty. Besides quantitative analysis, we examine common failure patterns in dynamic concurrency (\researchq{3}) and the mechanisms and conditions of observed benefits (\researchq{4}). By manually annotating complete execution trajectories (Cohen's $\kappa=0.77$~\cite{cohen1960agreement}), we develop a taxonomy with four main categories covering task orchestration, execution governance, global context management, and shared state and merge. These categories comprise 13 subcategories and 28 specific failure patterns. Moreover, we also find that dynamic concurrency brings benefits when sub-agents can complete complementary work, explore alternative solutions, or
independently validate and correct implementation from the main agent.

In summary, we make the following contributions:

\begin{itemize}[leftmargin=*]
  \item \textbf{Systematic Empirical Study.} We systematically evaluate dynamic concurrency in  frontier coding agents across long-horizon coding tasks, analyzing its effectiveness and  efficiency, along with its sub-agent delegation and coordination behaviors.
  \item \textbf{In-depth Failure Analysis.} We analyze paired concurrent and sequential trajectories and construct a comprehensive taxonomy of failure patterns in dynamic concurrency.
  \item \textbf{Large-Scale Dataset.} We release all 2,124 agent trajectories on long-horizon coding tasks with and without dynamic concurrency; moreover, the failure pattern annotations and analysis scripts are also released to support reproducible research on dynamic concurrency in coding agents.
\end{itemize}

\FloatBarrier
\section{Background and Related Work}\label{sec:background}

\parabf{Concurrency in Coding Agents.} Earlier coding agents organize agent collaboration through predefined roles and workflows, task graphs, or interaction protocols. For example, ChatDev and MetaGPT exemplify specialized roles and development workflows~\cite{qian2024chatdev,hong2023metagpt,huang2024agentcoder,islam2024mapcoder}; CodeR coordinates issue resolution through predefined task graphs~\cite{chen2024coder}. Multiagent Debate and SWE-Debate use repeated discussion rounds~\cite{du2024multiagentdebate,smit2024mad,li2026swedebate}, while Mixture-of-Agents uses layered aggregation~\cite{wang2024mixtureofagents}.

Different from the static paradigm, which orchestrates multiple sub-agents according to a predefined and fixed workflow, recent frontier coding agents propose \emph{dynamic concurrency} for multi-agent collaboration. In this paradigm, the main coding agent dynamically decides during execution whether to spawn sub-agents, which tasks to delegate, and how to coordinate and integrate their results. In this work, we study the native dynamic-concurrency mechanisms of Codex Multiagent Version 2, Claude Code Dynamic Workflow, and Kimi Code AgentSwarm~\cite{openaiCodexSubagents,anthropicDynamicWorkflows,moonshotKimiCode}.

\parabf{Empirical Studies of Coding Agents.}
To date, there have been extensive studies on coding agents. For example, several studies analyze  coordination architectures and debate settings of coding agents, examining how interaction design affects accuracy, cost, and latency~\cite{tian2026enagent,chun2026codingdebate,li2026swedebate}.
In addition, a substantial body of work has investigated failure modes in agents~\cite{lu2025autonomousfailures}. For example, TRAJEVAL exemplifies trajectory analyses of coding-agent errors and execution patterns associated with success or failure~\cite{liu2025automatedfailures,chen2026beyondfinalcode,bouzenia2025understanding,meng2026bugfixing,kim2026coherencecollapse,majgaonkar2025agentbehaviour,ma2026samesignal,ceka2026traceability}. Coordination studies, including MAST and AgentFail, examine inter-agent misalignment, information loss, and failure propagation~\cite{cemri2025multiagentfailures,lyu2026plannercoder,ma2026agentfail,seyedghorban2026observability}. TrajAudit and CodeTracer exemplify automated diagnosis of coding-agent trajectories~\cite{wang2026trajaudit,li2026codetracer}, complementing broader work on failure attribution and feedback-based recovery~\cite{zhu2025agentdebug,zhang2025failureattribution,ge2026famas,deshpande2025trail,zhang2026agentracer,lyu2026cam}. Different from existing work, our study focuses on dynamic concurrency, an emerging mechanism in frontier coding agents, and systematically investigates its behavior on challenging long-horizon coding tasks. In particular, we evaluate its effectiveness and efficiency at scale and characterize the failure patterns that arise when multiple sub-agents are coordinated via dynamic concurrency.

\begin{figure}[htbp]
  \centering
  \includegraphics[width=\textwidth,pagebox=cropbox]{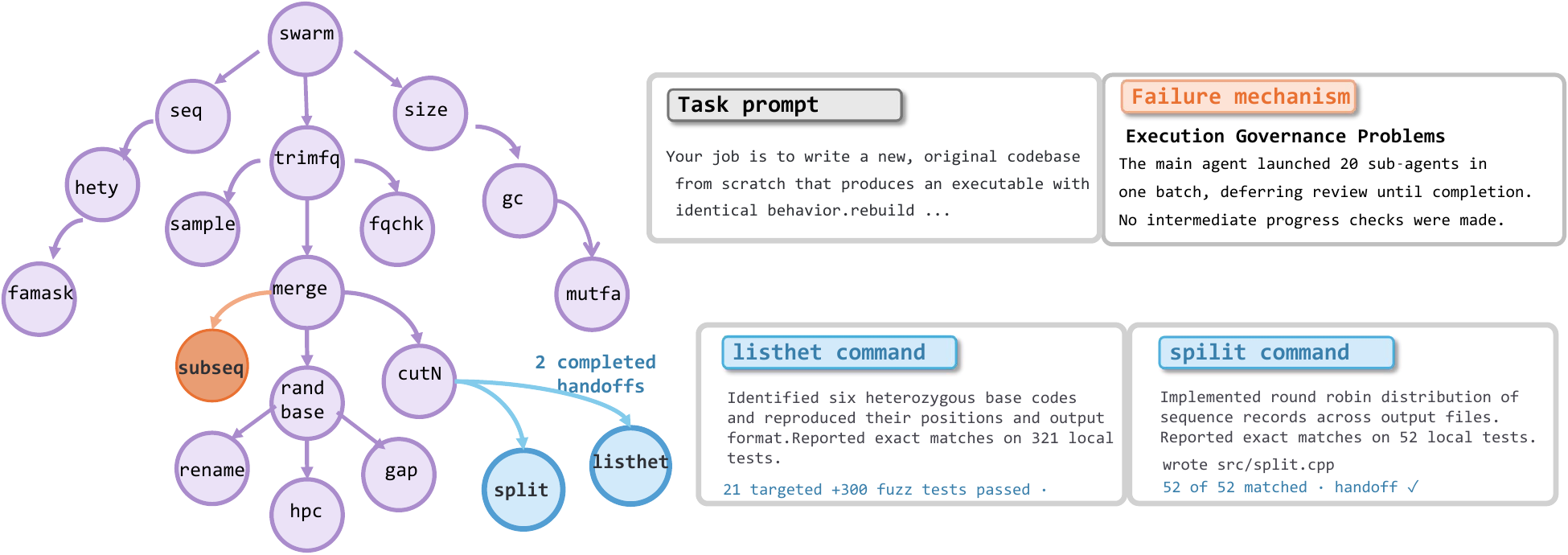}
  \caption{Dynamic concurrency in Kimi Code AgentSwarm on the \texttt{seqtk} reconstruction task.}
  \label{fig:illustrative-example-seqtk}
\end{figure}

\parabf{Illustrative Example on Dynamic Concurrency: Rebuilding seqtk.}
 Figure~\ref{fig:illustrative-example-seqtk} presents the execution graph of sub-agents when Kimi Code AgentSwarm is solving one long-horizon task  with dynamic concurrency (i.e., \texttt{lh3\_\_seqtk.94e7070} in ProgramBench). The goal of the task is to rebuild \texttt{seqtk}, a command line tool for processing biological sequences, using a reference executable and documentation. In Figure~\ref{fig:illustrative-example-seqtk}, each node represents a sub-agent, except the root node which is the main agent. The main agent first probes the reference and writes a FASTA/FASTQ sequence reader, a command dispatcher, and a build script. These components provide the common code used by the individual commands.

The process is proceeded with the following stages.
(i) At 17.5 minutes into the 40 minute run, the main agent assigns 22 subcommands to 20 sub-agents in one batch. These include sequence transformation (\texttt{seq}), statistics (\texttt{size} and \texttt{comp}), and sampling (\texttt{sample}). Each sub-agent receives specific source files and the shared interface, with instructions to implement its commands and compare their outputs and exit status against the reference. The main agent remains responsible for the common code. All 20 sub-agents start within 11 seconds and work in parallel. (ii) The main agent then waits for the entire batch to return. It assumes a sub-agent timeout of two hours, exceeding the 40 minute task budget. Although the \texttt{listhet} and \texttt{split} sub-agents report completion at minutes 22.3 and 26.7, the main agent makes no further tool calls and never reaches its planned integration and final tests. The other 18 sub-agents are aborted at the deadline. Five have already written command source files to the shared workspace, despite not reporting completion. (iii) The submitted program builds successfully, but some commands are missing. The sub-agent assigned \texttt{seq} never writes its implementation, so the official test invoking \texttt{seq -Q 64} receives ``unrecognized command'' instead of processed sequences. The program passes 92 of 440 tests. The command assignments allow implementation to proceed in parallel, but waiting for the entire batch leaves no time to check the available code, address missing commands, or test the combined program.

\FloatBarrier
\section{Methodology}
\label{sec:methodology}

\subsection{Research Questions}
\begin{itemize}[leftmargin=10pt]
\item \textbf{RQ1: Effectiveness and Efficiency.}
How does enabling dynamic concurrency affect task outcomes and resource use across long-horizon coding tasks?
\item \textbf{RQ2: Delegation and Coordination Behavior.} How do frontier coding agents organize and coordinate concurrent work across sub-agents over time?
\item \textbf{RQ3: Taxonomy of Concurrency Failure Patterns.} What failure patterns arise in frontier coding agents using dynamic concurrency?

\item \textbf{RQ4: Benefit Analysis of Dynamic Concurrency.} Through what mechanisms and under what coordination conditions, does dynamic concurrency improve effectiveness or efficiency?
\end{itemize}

\subsection{Studied Agents}
We select three representative frontier coding agents that support native dynamic concurrency. As reinforcement learning during agentic post-training may adapt a model to its scaffold's tool interfaces and interaction conventions~\cite{kimi2026k25}, substituting another model could disrupt this compatibility. We use agent--model configurations selected in July 2026, pairing Codex with GPT-5.4~\cite{openaiGPT54}, Claude Code with Claude Opus~5~\cite{anthropicModelConfiguration}, and Kimi Code with Kimi~K3~\cite{moonshotKimiK3}. Figure~\ref{fig:study-overview} illustrates their delegation and integration mechanisms. We use the native scaffold of all agents and request the highest reasoning effort.

\subsection{Benchmarks}
We mainly focus on long-horizon coding tasks as there is limited necessity for dispatching multiple sub-agents for simple and bounded tasks. To this end, we select four widely-used benchmarks for long-horizon software artifact construction and sustained software development, including (i) \textit{RepoZero}, which include Py2JS tasks (i.e., reconstructing Python packages in JavaScript) and C2Rust tasks (i.e., reconstructing C interfaces in Rust)~\cite{zhang2026repozero}; (ii) \textit{ProgramBench}, which requires rebuilding complete command line programs from reference executables and documentation~\cite{yang2026programbench}; (iii) \textit{NL2Repo Bench}, which requires constructing repositories from natural language specifications and an empty workspace~\cite{ding2026nl2repobench}; and (iv) \textit{LoopsBench}, which covers sustained work across dependent development units~\cite{li2026loopsbench}.
In addition, for comparison, we also include the classic coding benchmark \textit{SWE-bench Verified}~\cite{jimenez2024swebench,openai2024swebenchverified}, which requires to resolve one specific issue in an existing repository, as the representative of  bounded and short-horizon coding tasks.

Table~\ref{tab:benchmarks} summarizes the statistics of studied tasks. The ``Pop.'' column refers to the number of tasks in the full benchmark. However, executing the entire long-horizon benchmarks with diverse settings can be extremely expensive (e.g., running the entire LoopsBench once can cost \$11,354) given the massive tokens consumed for long horizon tasks. Therefore, for each benchmark, we make a representative sample of tasks across different difficulty groups (i.e., easy, medium, and hard). We adopt the original difficulty groups defined by each benchmark themselves. To ensure the sampled tasks still preserve each benchmark's difficulty distribution, we use stratified random sampling for each difficulty group: we select 25\% of tasks in each studied long-horizon benchmarks (i.e., RepoZero, ProgramBench, NL2Repo, and LoopsBench) and 20\% of the tasks in the comparison benchmark SWE-bench Verified. The detailed sample distribution is presented in Table~\ref{tab:benchmarks}.

Figure~\ref{fig:benchmark-implementation-size} further shows the implementation sizes (i.e., the lines of code of ground-truth outputs for each task) across different benchmarks and difficulty groups. Overall, the four studied long-horizon benchmarks require generating significantly larger scale of code than short-horizon tasks in SWE-bench Verified. In particular, LoopsBench stands as the even most complicated tasks (e.g., involving generating 26.8K lines of code) among all long-horizon benchmarks. In summary, our experiments include 354 distinct tasks, which are each executed with three coding agents against two settings (i.e., enable or disable dynamic concurrency), yielding 2,124 execution cells in total, with recorded token usage corresponding to an API cost of \$20,780.02 at standard rates.

\begin{table}[!htbp]
  \noindent\begin{minipage}[c]{0.535\linewidth}
  \caption{Task distribution across difficulty groups.}
  \label{tab:benchmarks}
  \raggedright
  \scriptsize
  \setlength{\tabcolsep}{2pt}
  \renewcommand{\arraystretch}{1.35}
  \begin{tabular*}{\linewidth}{@{}l@{\extracolsep{\fill}}rrrrrr@{}}
    \toprule
    Benchmark & Pop. & Sample & Easy & Med. & Hard & Cells \\
    \midrule
    \rowcolor[HTML]{EAF3FF}\multicolumn{7}{@{}l@{}}{\textbf{Short-horizon tasks}} \\*
    SWE Verified & 500 & 100 & 39 & 52 & 9 & 600 \\
    \midrule
    \rowcolor[HTML]{EAF3FF}\multicolumn{7}{@{}l@{}}{\textbf{Long-horizon tasks}} \\*
    Py2JS & 400 & 100 & 33 & 32 & 35 & 600 \\
    C2Rust & 200 & 50 & 15 & 18 & 17 & 300 \\
    NL2Repo & 104 & 26 & 6 & 12 & 8 & 156 \\
    ProgramBench & 200 & 50 & 7 & 36 & 7 & 300 \\
    LoopsBench & 112 & 28 & 4 & 11 & 13 & 168 \\
    \midrule
    Total & 1,516 & 354 & 104 & 161 & 89 & 2,124 \\
    \bottomrule
  \end{tabular*}
  \end{minipage}\hfill
  \begin{minipage}[c]{0.445\linewidth}
    \captionsetup{type=figure}
    \centering
    \includegraphics[width=\linewidth]{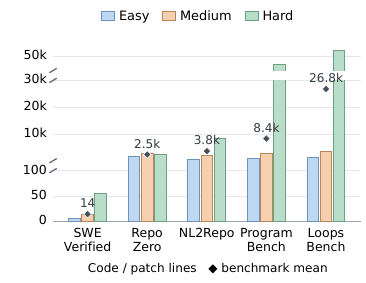}
    \caption{Implementation size distribution.}
    \label{fig:benchmark-implementation-size}
  \end{minipage}
\end{table}

\subsection{Experimental Process} \label{sec:experimental-design}
Each task is evaluated with all three coding agents in both modes, i.e., (i) concurrent mode which enables the dynamic concurrency, and (ii) sequential mode which disables the dynamic concurrency. For a fair comparison, we control key experimental conditions between two modes, including the backbone model, prompt, environment, and time budget. In particular, for each studied agent, we enable the dynamic concurrency with their native interface and default settings. Codex enables concurrency through \texttt{multi\_agent} and \texttt{multi\_agent\_v2}. Its sequential mode disables both options. Claude Code uses \texttt{ultracode} to activate Dynamic Workflow. Its sequential mode disables Workflow, Task, and Agent tools. Kimi Code enables AgentSwarm with \texttt{swarm\_mode=true} and uses \texttt{swarm\_mode=false} in sequential mode. In particular, for all studied agents, disabling dynamic concurrency guarantees that no sub-agents will be created. After execution, we collect the complete trajectories of both the main agent and all sub-agents, the submitted artifacts, and official test evaluator results, to facilitate our further analysis.

\section{Effectiveness and Efficiency (RQ1)}
\label{sec:rq1}
This RQ investigates how dynamic concurrency affects the overall effectiveness and efficiency of coding agents in long-horizon coding tasks. Table~\ref{tab:rq1-overview} presents overall results of three coding agents across five benchmarks. The column ``Conc.'' refers to the concurrent mode (i.e., enable dynamic concurrency) and the column ``Seq.'' refers to the sequential mode (i.e., disable dynamic concurrency). The rows ``Task pass'' and ``Test pass'' report the percentage of fully resolved tasks and the mean percentage of passing tests per task, respectively.

\subsection{Effectiveness}

\begin{table}[!htbp]
  \caption{Official task quality and resource use by benchmark, agent, and execution mode. Arrows indicate whether higher ($\uparrow$) or lower ($\downarrow$) is better. Bold marks the better result within each agent and metric.}
  \label{tab:rq1-overview}
  \centering
  \footnotesize
  \setlength{\tabcolsep}{3.0pt}
  \renewcommand{\arraystretch}{1.04}
  \begin{tabular*}{\textwidth}{@{}l@{\extracolsep{\fill}}lrrrrrr@{}}
    \toprule
    & & \multicolumn{2}{c}{Codex} & \multicolumn{2}{c}{Claude Code} & \multicolumn{2}{c}{Kimi Code} \\
    \cmidrule(lr){3-4}\cmidrule(lr){5-6}\cmidrule(l){7-8}
    Benchmark & Metric & Conc. & Seq. & Conc. & Seq. & Conc. & Seq. \\
    \midrule
    \rowcolor[HTML]{EAF3FF}\multicolumn{8}{@{}l@{}}{\textbf{Short-horizon tasks}} \\*
    \multirow{5}{*}{SWE-Verified} & Shell calls $\downarrow$ & 115.8 & \textbf{53.1} & 21.3 & \textbf{19.3} & 23.8 & \textbf{19.9} \\*
     & Time (min) $\downarrow$ & 5.1 & \textbf{4.5} & 27.8 & \textbf{17.5} & 12.3 & \textbf{11.5} \\*
     & Tokens (M) $\downarrow$ & 3.37 & \textbf{1.34} & 2.14 & \textbf{1.02} & 1.01 & \textbf{0.98} \\*
     & Task pass (\%) $\uparrow$ & 71.0 & 71.0 & 59.0 & \textbf{83.0} & 59.0 & \textbf{77.0} \\*
     & Test pass (\%) $\uparrow$ & 89.0 & \textbf{89.6} & 78.7 & \textbf{90.6} & 78.1 & \textbf{93.4} \\
    \midrule
    \rowcolor[HTML]{EAF3FF}\multicolumn{8}{@{}l@{}}{\textbf{Long-horizon tasks}} \\*
    \multirow{5}{*}{RepoZero} & Shell calls $\downarrow$ & 135.9 & \textbf{60.4} & 232.1 & \textbf{23.0} & 41.2 & \textbf{15.7} \\*
     & Time (min) $\downarrow$ & 7.2 & \textbf{6.3} & 39.3 & \textbf{29.0} & 19.3 & \textbf{17.1} \\*
     & Tokens (M) $\downarrow$ & 2.47 & \textbf{0.89} & 4.03 & \textbf{2.42} & 1.33 & \textbf{0.82} \\*
     & Task pass (\%) $\uparrow$ & \textbf{40.7} & 40.0 & 26.7 & \textbf{42.7} & \textbf{38.7} & 37.3 \\*
     & Test pass (\%) $\uparrow$ & 70.9 & \textbf{71.2} & 41.5 & \textbf{64.9} & 65.5 & \textbf{66.7} \\
    \midrule
    \multirow{5}{*}{NL2Repo} & Shell calls $\downarrow$ & 155.4 & \textbf{47.3} & 334.6 & \textbf{35.2} & 45.8 & \textbf{23.6} \\*
     & Time (min) $\downarrow$ & 9.9 & \textbf{9.4} & 39.0 & \textbf{34.3} & 24.4 & \textbf{18.4} \\*
     & Tokens (M) $\downarrow$ & 8.00 & \textbf{2.87} & 8.55 & \textbf{6.53} & 2.70 & \textbf{2.21} \\*
     & Task pass (\%) $\uparrow$ & \textbf{57.7} & 34.6 & \textbf{65.4} & 57.7 & 46.2 & \textbf{57.7} \\*
     & Test pass (\%) $\uparrow$ & \textbf{88.8} & 86.2 & \textbf{92.2} & 88.4 & 81.3 & \textbf{87.9} \\
    \midrule
    \multirow{5}{*}{ProgramBench} & Shell calls $\downarrow$ & 496.7 & \textbf{180.4} & 608.2 & \textbf{68.3} & 317.3 & \textbf{132.5} \\*
     & Time (min) $\downarrow$ & 24.9 & \textbf{23.4} & 38.9 & \textbf{38.7} & \textbf{39.9} & 40.1 \\*
     & Tokens (M) $\downarrow$ & 27.05 & \textbf{9.19} & \textbf{6.02} & 6.18 & 11.64 & \textbf{6.05} \\*
     & Task pass (\%) $\uparrow$ & 0.0 & 0.0 & 0.0 & 0.0 & 0.0 & 0.0 \\*
     & Test pass (\%) $\uparrow$ & \textbf{58.3} & 53.5 & 0.5 & \textbf{1.8} & \textbf{1.7} & 1.5 \\
    \midrule
    \multirow{5}{*}{LoopsBench} & Shell calls $\downarrow$ & 505.6 & \textbf{144.6} & 791.6 & \textbf{107.4} & 618.0 & \textbf{120.8} \\*
     & Time (min) $\downarrow$ & 18.8 & \textbf{17.7} & 91.5 & \textbf{74.8} & 55.8 & \textbf{52.7} \\*
     & Tokens (M) $\downarrow$ & 51.65 & \textbf{11.45} & 21.73 & \textbf{15.52} & 41.49 & \textbf{11.29} \\*
     & Task pass (\%) $\uparrow$ & \textbf{25.0} & 21.4 & \textbf{21.4} & 7.1 & 25.0 & 25.0 \\*
     & Test pass (\%) $\uparrow$ & \textbf{58.1} & 52.1 & \textbf{48.7} & 45.8 & \textbf{66.6} & 62.5 \\
    \bottomrule
  \end{tabular*}
\end{table}

\noindent\textbf{Pass Rate Improvement.}
Dynamic concurrency shows promise in improving task and test pass rates in several long-horizon tasks, with the broadest gains on the most long-horizon benchmark LoopsBench.
First, on LoopsBench, which contains the longest-horizon tasks among all studied benchmarks, dynamic concurrency improves task pass rates for Codex and Claude Code by up to 14.3 percentage points, while Kimi Code's task pass rate remains unchanged. Test pass rates improve for all three agents, with gains of up to 6.0 percentage points.
Second, the benefits are more mixed on the other long-horizon benchmarks. On NL2Repo, dynamic concurrency improves the task pass rates of Codex and Claude Code but decreases that of Kimi Code. On ProgramBench, which remains particularly challenging for all studied frontier coding agents (with a task pass rate of 0\% across all agents), concurrency nevertheless improves partial progress, as measured by test pass rate, for Codex and Kimi Code. Substantial variation can arise even within the same benchmark. For example, Codex's nearly unchanged aggregate task pass rate on RepoZero combines a decline on Py2JS (2.0/1.7 percentage-point decreases in task/test pass rates) with an improvement on C2Rust (6.0/2.4 percentage-point increases in task/test pass rates). The loss of two resolved Py2JS tasks and gain of three resolved C2Rust tasks largely offset each other, leaving just one additional resolved task out of 150 (40.0\% to 40.7\% overall).
Third, dynamic concurrency provides no improvement on the bounded, short-horizon tasks in SWE-bench Verified. Moreover, Codex remains unchanged, while Claude Code and Kimi Code exhibit substantial decreases in pass rate. For Claude Code, its task pass rate decreases by 24.0 percentage points on SWE-bench Verified but increases by 14.3 percentage points on LoopsBench.  In particular, pooling both task groups, we conduct exact McNemar tests~\cite{mcnemar1947correlated,fagerland2013mcnemar} with Holm correction~\cite{holm1979multiple} across the three agents and assess significance by comparing the adjusted $p$-values against $\alpha=0.05$. The overall declines for Claude Code ($p=5.08348\times10^{-7}$) and Kimi Code ($p=0.0132176$) are statistically significant, whereas the improvement for Codex ($p=0.229481$) is not.

We further compare average test pass rates across the four long-horizon benchmarks, ordered by their mean implementation size (Figure~\ref{fig:benchmark-implementation-size}). This measure also captures progress on tasks that are not fully solved. Averaged equally across the three agents, the difference between concurrent and sequential execution rises from $-8.29$ percentage points on RepoZero to $-0.04$ on NL2Repo, $+1.24$ on ProgramBench, and $+4.31$ on LoopsBench. The effectiveness difference across these benchmarks indicates that larger expected implementation sizes are associated with a greater average test pass advantage from dynamic concurrency.

\begin{figure}[htbp]
  \centering
  \includegraphics[width=\textwidth]{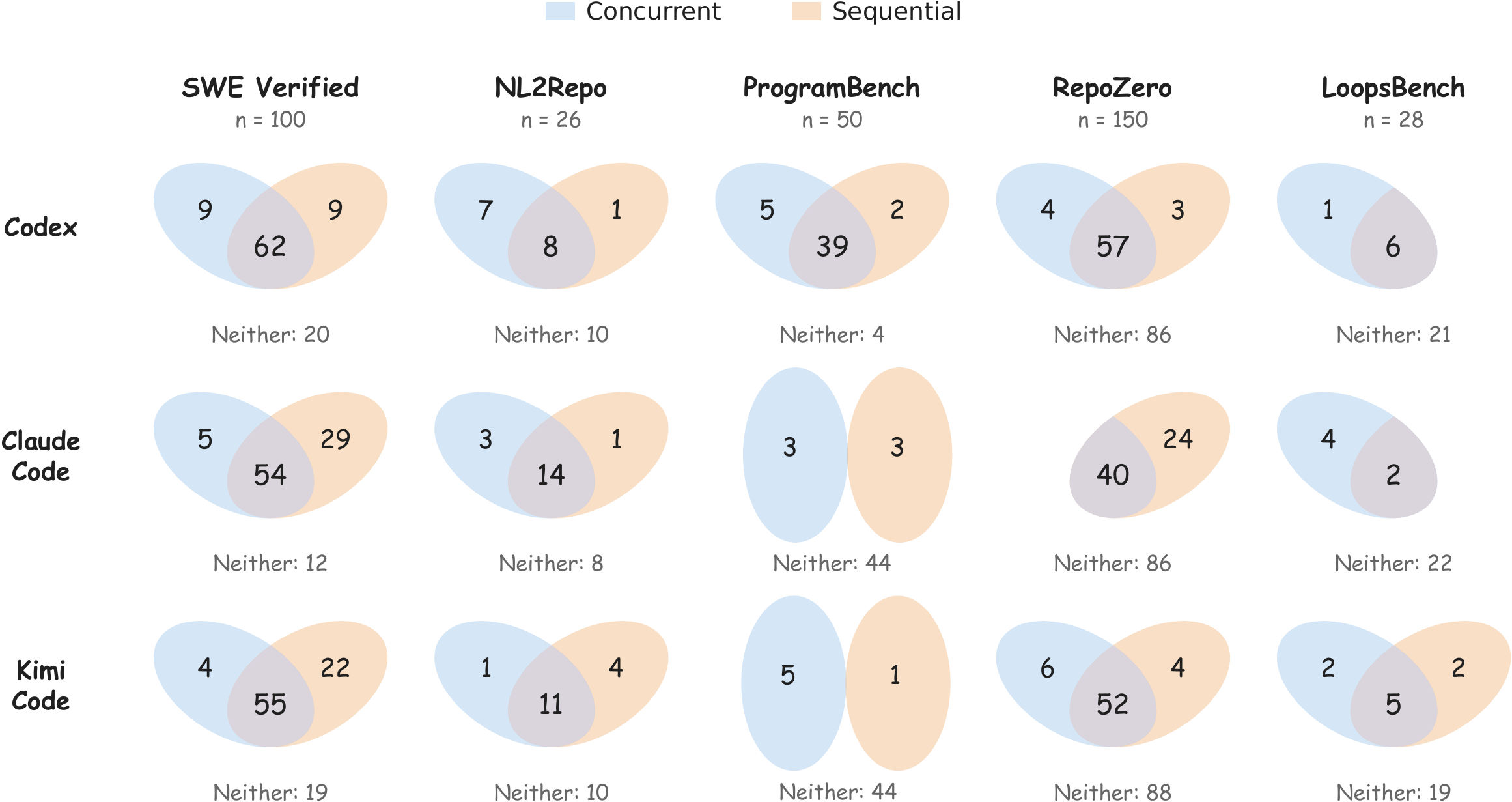}
  \caption{Overlap of task outcomes under concurrent and sequential execution by agent and benchmark.}
  \label{fig:rq1-resolution-overlap}
\end{figure}

\noindent\textbf{Overlap Analysis.} Numeric pass rates alone do not reveal whether concurrent and sequential execution succeed on the same tasks. We therefore further examine the overlap of successful tasks between the two modes for each agent in Figure~\ref{fig:rq1-resolution-overlap}. In particular, for ProgramBench, which requires reconstructing complete programs from documentation and black-box access to reference executables~\cite{yang2026programbench}, no sampled task is fully resolved by any agent in either mode. We therefore consider a ProgramBench task successful if it achieves a test pass rate above zero, while using full task resolution for the other benchmarks. Counts below the schematic ellipses denote tasks satisfying neither criterion, and empty regions are omitted.

Overall, the dynamic concurrency consistently enables unique successes that are not achieved under sequential execution. In particular, we can observe unique resolved tasks from the concurrent model almost across all agents and benchmarks.  In the most pronounced case, Claude Code solves four LoopsBench tasks exclusively under concurrency, with no task uniquely solved under sequential execution. Notably, concurrency can expose complementary capabilities even when it does not improve aggregate performance. For example, on the bounded and short horizon tasks in SWE-bench Verified, Claude Code still uniquely solves five tasks under concurrency despite its lower overall pass rate, while Codex achieves identical aggregate performance across the two modes but uniquely solves nine tasks in each mode. These results suggest that concurrency can expand the capabilities of coding agents within the sequential mode. We further examine concurrency failure patterns in RQ3 and the mechanisms and conditions of observed benefits in RQ4.

\begin{takeawaybox}
\textbf{Finding 1:} Dynamic concurrency offers a promising way to expand the capabilities of coding agents beyond sequential execution, particularly on long-horizon tasks. Larger expected implementation sizes are associated with greater average test pass gains across these benchmarks.
\end{takeawaybox}

\subsection{Resource Efficiency}
As shown in Table~\ref{tab:rq1-overview},  the column ``Shell calls'' includes the number of calls made by both the main agent and all sub-agents;
the columns ``Time'' and ``Token'' present the agent execution time and overall costs based on recorded provider usage.

\noindent\textbf{Resource Efficiency.}
Overall, dynamic concurrency incurs substantially higher computational cost than sequential execution. As shown in Table~\ref{tab:rq1-overview}, mean shell calls increase under concurrency for every agent and benchmark, often by several fold, reflecting the additional exploration and execution performed by dispatching multiple sub-agents. Token consumption similarly increases, reaching 1.41--3.31$\times$ the sequential level across agents. More importantly, these additional resources do not translate proportionally into successful outcomes. When normalizing total token consumption across both successful and failed runs by the number of successfully resolved tasks, token usage per solved task increases for all three agents. Even Codex, the  agent showing a small overall gain in resolved tasks, consumes 24.12 million recorded tokens per solved task under concurrency, compared with 7.69 million under sequential execution. These results indicate that the current concurrency mechanisms generally trade substantially greater resource consumption for limited aggregate effectiveness gains.

\noindent\textbf{Execution Time.}
Interestingly, despite parallel execution, dynamic concurrency does not generally reduce end-to-end runtime. Mean runtime increases for all three agents overall and in 14 of the 15 agent-benchmark combinations in Table~\ref{tab:rq1-overview}. For example, Claude Code takes 27.8 minutes per SWE-bench Verified task under concurrency, compared with 17.5 minutes under sequential execution, despite resolving fewer tasks. Thus, the parallelism introduced by dynamic concurrency does not automatically translate into lower end-to-end latency. We will perform a qualitative analysis in RQ3 to investigate the root causes for additional latency from concurrent mode.

The costs and time in Table~\ref{tab:rq1-overview} are calculated based on average over both successful and unsuccessful tasks. We further analyze the costs of the tasks that are successfully solved by both concurrent and sequential modes. Of the 366 cases solved in both modes, concurrent execution finishes faster in 102 cases (27.9\%). In particular, in 46 cases (12.6\%), concurrent mode completes the task substantially faster than sequential mode (at least 20\% less time). The observation implies that meaningful speedups can be achieved from the concurrent mode when proper coordination and execution are scheduled. We will further perform a qualitative analysis in RQ4 to investigate the conditions underlying these successful speedup.

\begin{takeawaybox}
\textbf{Finding 2:} Dynamic concurrency incurs substantially higher resource consumption and runtime on average, with limited returns in effectiveness. Nevertheless, it shows potential in achieving meaningful speedup (e.g., on 12.6\% of tasks solved in both modes), highlighting opportunities for more selective and effective concurrency.
\end{takeawaybox}

\FloatBarrier
\section{Delegation and Coordination Behavior (RQ2)}\label{sec:rq2}
While RQ1 focuses on outcome-centric analysis, this RQ focuses on process-centric analysis by analyzing the delegation and coordination behavior during agent execution, i.e., how sub-agents are dynamically dispatched and scheduled when the long-horizon tasks proceed.

\subsection{Delegation Scale}\label{sec:rq2:scale}
\begin{figure}[!htbp]
  \centering
  \includegraphics[width=\textwidth]{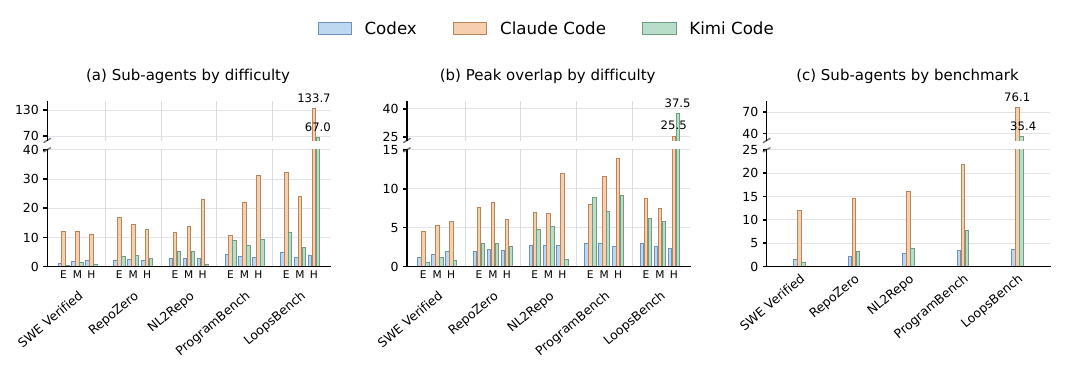}
  \caption{Mean delegation counts and peak sub-agent overlap by benchmark and task difficulty.}
  \label{fig:rq2-delegation-means}
\end{figure}

We analyze the trajectories of all the 354 concurrent executions for each coding agent. Each sub-agent context is counted once per execution. Based on the creation and termination timestamps of each sub-agent, we consider two sub-agents to overlap if their available time windows intersect.
Figure~\ref{fig:rq2-delegation-means} compares mean counts of created sub-agents (a), peak counts of overlapped sub-agents (b), and mean sub-agent counts for each benchmark (c). E, M, and H denote easy, medium, and hard tasks. Created counts accumulate across an execution, whereas the peak captures how many sub-agents coexist. Repeated waves of delegation can therefore increase the total without increasing the peak.

\noindent\textbf{Delegation scale across task difficulty.}
Overall, coding agents tend to delegate more sub-agents on harder tasks. Across the pooled sample, weighting each execution equally, the mean number of created sub-agents increases from 6.69 on easy tasks to 7.53 on medium tasks and 15.84 on hard tasks, while the mean peak number of concurrently active sub-agents increases from 3.66 to 4.61 and 6.70, respectively.
Delegation increases most from easy to hard tasks on LoopsBench, which has the longest task horizons. Averaged across the three agents, the mean number of created sub-agents rises from 16.25 to 68.18 per task, while the mean peak rises from 6.00 to 21.77. LoopsBench accounts for 14.6\% of the hard tasks but 62.9\% of all sub-agents created on hard tasks. Similarly, for Claude Code on ProgramBench, the mean number of created sub-agents increases from 10.71 to 22.08 and 31.29 across easy, medium, and hard tasks, with corresponding mean peaks of 8.00, 11.56, and 13.86. However, there are exceptions such as for bounded tasks from SWE-bench Verified. For another example,  Claude Code does not necessarily create or maintain more sub-agents on harder tasks from RepoZero. We further test the association between task difficulty and each delegation metric using two-sided Spearman permutation tests with Holm correction~\cite{holm1979multiple} across all 30 comparisons. The number of created sub-agents increases significantly with difficulty for Claude Code on ProgramBench ($\rho = 0.470$, adjusted $p = 0.015$) and Codex on SWE-bench Verified ($\rho = 0.316$, adjusted $p = 0.042$). The association is significant in some settings, but task structure and coordination mechanisms also shape delegation. A task can become harder because one core component requires more effort or because other components must wait for it. This can leave the number of independent subtasks unchanged even as the total work increases.

\noindent\textbf{Delegation preference across agents.}
Codex generally maintains small groups of sub-agents, with mean peaks ranging from 1.21 to 3.00 across task groups. On hard LoopsBench tasks, Claude Code creates 133.69 sub-agents per task on average, with a mean peak of 25.46. Kimi Code creates fewer sub-agents, averaging 67.00, but reaches a higher mean peak of 37.46. Keeping part of the implementation with the main agent allows Codex to use fewer sub-agents. Claude Code can create new groups in successive workflow stages, increasing the total count without keeping all of them active at once. Kimi Code can launch a large batch together, producing a higher peak despite a smaller total count.

\begin{takeawaybox}
\textbf{Finding 3:} Coding agents generally scale up sub-agent delegation as task difficulty increases on long-horizon tasks. Different coding agents have different preferences in delegation scale when enabled with dynamic concurrency, e.g., Codex often creates and maintains fewer sub-agents.
\end{takeawaybox}

\FloatBarrier

\subsection{Temporal Dynamics and Context Overlap}
To characterize when and for how long coding agents exploit concurrency, we normalize each execution trace from the first to the last recorded timestamp of the main agent or any sub-agent, including successive delegation rounds and the gaps between them.  We define the \emph{overlap fraction} as the proportion of the normalized trace during which at least two sub-agents are simultaneously active, excluding the main agent. Executions with no such overlap contribute an overlap fraction of zero. Thus, this metric captures concurrency among sub-agents rather than parallel activity between the main agent and a single sub-agent.

\begin{figure}[!htbp]
  \centering
  \includegraphics[width=\textwidth]{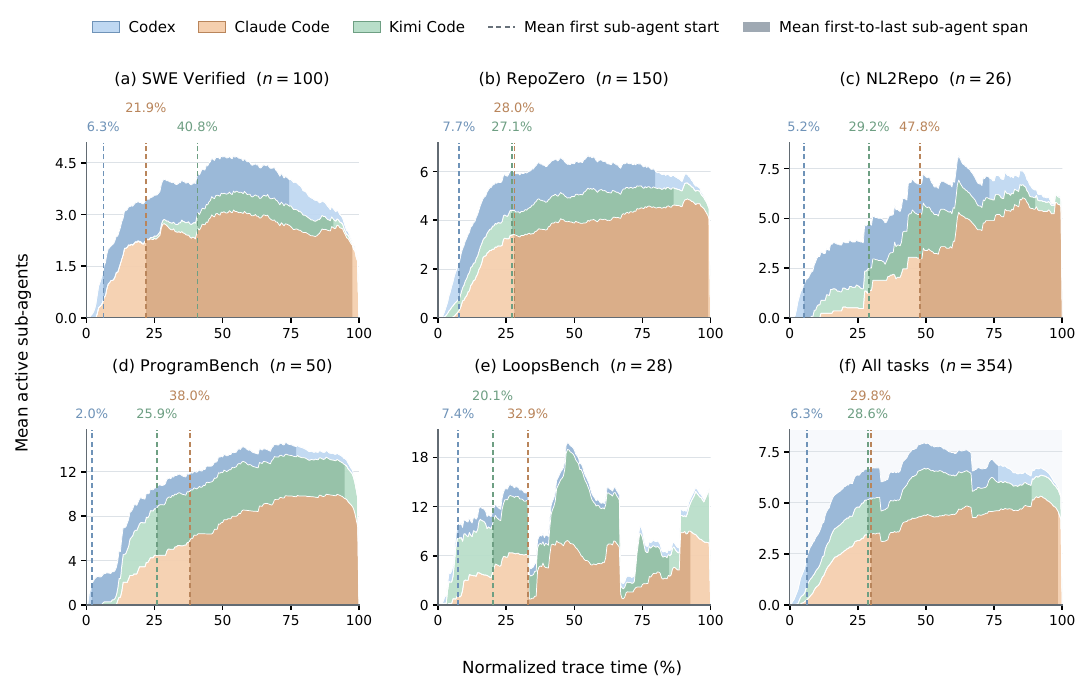}
  \caption{Mean observed sub-agent activity over normalized trace time.}
  \label{fig:rq2-timeline}
\end{figure}

In addition, Figure~\ref{fig:rq2-timeline} further illustrates the temporal differences through mean sub-agent activity profiles. Each band's thickness represents the mean number of active sub-agents for one coding agent, with the stacked bands aggregating separate executions. The overall panel weights tasks equally, and vertical scales differ across panels. Dashed lines indicate the mean first sub-agent start, while darker bands extend to the mean last observed endpoint among executions with sub-agents. These spans include any intervening gaps in sub-agent activity. The profiles nevertheless reveal distinct temporal structures, including differences in delegation onset, sustained activity, and successive waves of sub-agent creation.

\noindent\textbf{Delegation and Overlap Timing.}
The three coding agents exhibit substantially different temporal patterns in when they initiate delegation. Among executions that create sub-agents, Codex delegates substantially earlier, with the first sub-agent starting at a median of only 4.0\% of the trace, compared with 22.1\% for Claude Code and 26.0\% for Kimi Code. After inspecting Codex's implementation\cite{openaiCodex0145Source}, we find that Codex's spawn interface launches a child from a single task assignment and returns control to the main agent, allowing sub-agent work to begin before the rest of the task is fully planned. Preparing a workflow or batch involves organizing several assignments before launch: the archived Claude Code scripts specify agent calls and dependencies~\cite{concurrencyFailuresArtifact}, while Kimi Code's swarm instructions place exploration and decomposition before batch submission~\cite{moonshotKimi0292Source}.
In addition, a similar pattern emerges in when sub-agents begin to overlap in execution: based on our statistics, among executions exhibiting overlap, the first overlap occurs at a median of 4.7\%, 22.5\%, and 26.9\% of the trace for Codex, Claude Code, and Kimi Code, respectively. On the other hand, creating sub-agents does not necessarily result in concurrent execution. Among the 354 executions for each agent, 24.3\% of Codex, 1.7\% of Claude Code, and 2.8\% of Kimi Code though create sub-agents but their execution does not introduce any overlap among them. This distinction is particularly pronounced for Codex, i.e., despite delegating early and frequently creating sub-agents, many of its executions do not actually run multiple sub-agents concurrently.

\noindent\textbf{Overlap Duration.}
Beyond when concurrency begins, agents also differ substantially in how long they sustain overlapping sub-agent execution. Based on our statistics, the median overlap fraction across all tasks is 23.9\% for Codex, 52.7\% for Claude Code, and 15.6\% for Kimi Code. Claude Code therefore exhibits the most sustained concurrency: although it begins delegation later than Codex, multiple sub-agents overlap for more than half of the execution trace at the median. In contrast, Codex initiates delegation much earlier but sustains considerably less sub-agent overlap, while Kimi Code exhibits the lowest median overlap fraction. Such a temporal analysis further complement the peak-concurrency analysis in Section~\ref{sec:rq2:scale}, as similar peak counts of sub-agents can correspond to substantially different temporal behaviors depending on whether concurrency occurs briefly or persists throughout execution.

\noindent\textbf{Persistence of Delegation.}
The agents differ in how long sub-agent activity persists toward the end of execution. Among executions with sub-agents, the median last  sub-agent endpoint occurs at 83.3\% of the trace for Codex, compared with 100.0\% for Claude Code and 93.6\% for Kimi Code. Our inspection of Codex's code and trajectories shows that results are delivered individually and the main agent can resume when messages arrive~\cite{openaiCodex0145Source}, allowing implementation and checking to continue after delegated subtasks finish and leaving a longer final interval without sub-agent activity. The archived Claude Code workflow scripts create additional agents in later review stages, keeping delegation active as the workflow progresses toward completion~\cite{concurrencyFailuresArtifact}. Kimi Code's AgentSwarm implementation keeps the main agent waiting until the batch finishes~\cite{moonshotKimi0292Source}. These workflow stages and batch waits delay the completion of delegated tasks more closely to the end of execution.

\begin{takeawaybox}
\textbf{Finding 4:} Coding agents exhibit distinct temporal concurrency strategies. In particular, Codex delegates early but sustains relatively limited sub-agent overlap, whereas Claude Code delegates later but maintains the most sustained concurrency. Kimi Code maintains shorter concurrency with the lowest median overlap fraction.
\end{takeawaybox}

\FloatBarrier
\section{Taxonomy of Concurrency Failure Patterns (RQ3)}\label{sec:rq3}
This RQ manually investigates the common failure patterns in dynamic concurrency.

\subsection{Methodology}
We manually annotate all 1,062 execution trajectories generated by coding agents operating with dynamic concurrency. For each case, annotators review the complete concurrent trajectory alongside the corresponding sequential trajectory of the same coding agent on the same task, together with the execution outcomes and official test results from both runs. The review covers the main agent and all sub-agents (including nested ones), workflow and status records, and generated artifacts.

Two annotators first analyze 50 trajectories to establish an initial codebook through open coding~\cite{stol2016grounded}. They identify concrete coordination episodes and assign descriptive codes capturing observed behaviors and their execution- or artifact-level symptoms. From this analysis, we develop a taxonomy with three levels: \emph{main categories} describe broad areas of coordination failure, \emph{subcategories} distinguish forms of failure within each main category, and \emph{failure patterns} capture their concrete manifestations.

Three annotators participate in the formal annotation stage, with each case independently annotated by two of them using the codebook. For a case exhibiting concurrency failures, each annotator selects one or more applicable failure pattern labels and records the reasoning supporting the assignments. Cases with no identified failure pattern receive no label. Each failure pattern belongs to one subcategory within one main category, so the broader labels follow from the selected patterns without separate annotation decisions. We report a single inter-annotator agreement value for the independent failure pattern annotations (Cohen's $\kappa=0.77$)~\cite{cohen1960agreement}. After independent annotation, the annotators discuss disagreements, and a senior author adjudicates any unresolved disputes.

\subsection{Taxonomy}
The resulting taxonomy comprises four main categories, 13 subcategories, and 28 failure patterns, as shown in Table~\ref{tab:taxonomy}. These patterns occur in 650 reviewed concurrent execution trajectories, with 804 pattern instances in total. An execution trajectory may contain multiple failure patterns, and table counts and percentages are computed over the 804 instances. Each subcategory also has a failure type: \emph{functional failures} violate coordination requirements related to ownership, information exchange, isolation, observability, or acceptance, while \emph{performance failures} reflect ineffective coordination under the available execution budget. Examples of each failure pattern are available in our artifacts~\cite{concurrencyFailuresArtifact}.

\begin{table}[!tbp]
  \centering
  \scriptsize
  \caption{Taxonomy of concurrency failure patterns.}
  \label{tab:taxonomy}
  \begin{tblr}{
    width = \linewidth,
    colspec = {Q[l,wd=85.6pt] Q[l,wd=132.5pt] X[l] Q[l,wd=36.7pt]},
    cells = {valign=m},
    colsep = 2.5pt,
    rowsep = 2pt,
    hlines = {0.35pt,black},
    vlines = {0.35pt,black},
    hline{1,2,9,17,23,30} = {0.65pt,black},
    row{1} = {font=\bfseries},
    row{2-8} = {bg=orchestrationrow},
    row{9-16} = {bg=governancerow},
    row{17-22} = {bg=contextrow},
    row{23-29} = {bg=mergerow},
  }
    Main category & Subcategory & Failure pattern & Failure type \\
    \SetCell[r=7]{l,font=\bfseries} {A. Task Orchestration Problems \\ 229 (28.5\%)} & \SetCell[r=2]{l} Missing Owner \mbox{(16; 2.0\%)} & No Implementation Owner (14; 1.7\%) & \SetCell[r=2]{l} Functional \\
     &  & No Assembly Owner (2; 0.2\%) &  \\
     & \SetCell[r=2]{l} Pseudo-Concurrency \mbox{(47; 5.8\%)} & Preflight-Gated Work (7; 0.9\%) & \SetCell[r=2]{l} Performance \\
     &  & Serial Investigation (40; 5.0\%) &  \\
     & \SetCell[r=3]{l} Load Imbalance \mbox{(166; 20.6\%)} & Critical-Path Starvation (3; 0.4\%) & \SetCell[r=3]{l} Performance \\
     &  & Fan-out Budget Exhaustion (154; 19.2\%) &  \\
     &  & Oversized Sub-Agent Task (9; 1.1\%) &  \\
    \SetCell[r=8]{l,font=\bfseries} {B. Execution Governance Problems \\ 233 (29.0\%)} & \SetCell[r=5]{l} Wrong Start, Join, or Result Timing \mbox{(166; 20.6\%)} & Early Sub-Agent Termination (128; 15.9\%) & \SetCell[r=5]{l} Functional \\
     &  & Late Finalization (5; 0.6\%) &  \\
     &  & Merge after Verification (7; 0.9\%) &  \\
     &  & Missing Implementation Join (9; 1.1\%) &  \\
     &  & Missing Verifier Return (17; 2.1\%) &  \\
     & \SetCell[r=1]{l} Failure Propagation \mbox{(42; 5.2\%)} & No Failure Takeover (42; 5.2\%) & \SetCell[r=1]{l} Functional \\
     & \SetCell[r=2]{l} No Active Main Agent Monitoring \mbox{(25; 3.1\%)} & Unused Completed Result (13; 1.6\%) & \SetCell[r=2]{l} Performance \\
     &  & Blind Timeout Wait (12; 1.5\%) &  \\
    \SetCell[r=6]{l,font=\bfseries} {C. Global Context Management Problems \\ 75 (9.3\%)} & \SetCell[r=1]{l} Missing Task Requirements \mbox{(23; 2.9\%)} & Incomplete Sub-Agent Brief (23; 2.9\%) & \SetCell[r=1]{l} Functional \\
     & \SetCell[r=1]{l} Missing Interface Contract \mbox{(5; 0.6\%)} & Missing Cross-Agent Contract (5; 0.6\%) & \SetCell[r=1]{l} Functional \\
     & \SetCell[r=2]{l} Overloaded Handoff \mbox{(4; 0.5\%)} & Lossy Handoff (1; 0.1\%) & \SetCell[r=2]{l} Functional \\
     &  & Bulk Handoff Overload (3; 0.4\%) &  \\
     & \SetCell[r=1]{l} Unsupported Global Completion \mbox{(18; 2.2\%)} & Unverified Global Completion (18; 2.2\%) & \SetCell[r=1]{l} Functional \\
     & \SetCell[r=1]{l} Unchanged Retry \mbox{(25; 3.1\%)} & Checkpoint-Free Retry (25; 3.1\%) & \SetCell[r=1]{l} Performance \\
    \SetCell[r=7]{l,font=\bfseries} {D. Shared State and Merge Problems \\ 267 (33.2\%)} & \SetCell[r=6]{l} Concurrent Writes \mbox{(213; 26.5\%)} & Cross-File Scope Collision (5; 0.6\%) & \SetCell[r=6]{l} Functional \\
     &  & Same-File Collision (42; 5.2\%) &  \\
     &  & Source Overwrite (14; 1.7\%) &  \\
     &  & Deliverable Overwrite (120; 14.9\%) &  \\
     &  & Final-Tree Overwrite (6; 0.7\%) &  \\
     &  & Unisolated Workspace Writes (26; 3.2\%) &  \\
     & \SetCell[r=1]{l} Shared Environment Contamination \mbox{(54; 6.7\%)} & Artifact Leakage (54; 6.7\%) & \SetCell[r=1]{l} Functional \\
  \end{tblr}
\end{table}

\subsubsection*{A. Task Orchestration Problems}
This category arises when the main agent fails to effectively divide and allocate work among itself and sub-agents. Specifically, required work may be left unassigned (\emph{Missing Owner}), seemingly parallel execution may leave critical work sequential (\emph{Pseudo-Concurrency}), or uneven delegation may leave insufficient resources for critical work (\emph{Load Imbalance}). Task orchestration accounts for 229 (28.5\%) of all identified pattern instances.

\noindent\textbf{A1. Missing Owner.}
Some required work is never assigned to any agent, leaving part of the implementation or final assembly unfinished. In Kimi Code's NL2Repo verifiers reconstruction, five sub-agents are assigned to parsers and rubrics, utilities and scripts, documentation, example environments, and a mock client, but no agent is assigned the core \texttt{verifiers/envs} module. One sub-agent later reports that the missing environment classes prevent end-to-end testing, and the submitted package passes none  official tests. Here, the main agent divides the work but leaves a critical component without an owner.

\noindent\textbf{A2. Pseudo-Concurrency.}
Multiple sub-agents run in parallel, but the work that determines task completion remains sequential. For example, the main agent may parallelize investigation but wait for all investigation results before starting implementation. In Kimi Code's ProgramBench tailspin reconstruction, 17 sub-agents concurrently investigate different behaviors of the log highlighter, but none is assigned to implementation. The main agent waits for their reports before starting implementation itself, and the execution ends during implementation planning without producing a working replacement, passing 0 of 785 tests. Thus, although many sub-agents run concurrently, the concurrency does not accelerate the critical implementation work.

\noindent\textbf{A3. Load Imbalance.}
This is the most frequent subcategory within task orchestration, accounting for 20.6\% of all pattern instances and 39.8\% of those identified for Claude Code. It occurs when too much effort is spent on some parallel activities, leaving insufficient time or resources for critical work, i.e.,  when the main agent launches too many sub-agents or assigns too much work to individual sub-agents. In Claude Code's ProgramBench tailspin reconstruction, the main agent starts building a Rust scaffold while a workflow creates sub-agents to investigate ten behavior areas and launches additional critics. These investigations and repeated delegations continue until the deadline, when unfinished sub-agents are terminated without returning their results. The main agent is left with an incomplete crate that cannot build. Here, excessive investigation and delegation consume the execution budget needed to complete the implementation.

\subsubsection*{B. Execution Governance Problems}
Execution governance problems arise when the main agent fails to properly manage sub-agents after delegation. Specifically, sub-agent work may be handled at the wrong time (\emph{Wrong Start, Join, or Result Timing}), failed work may be left unrecovered (\emph{Failure Propagation}), or the main agent may fail to monitor progress and intervene when needed (\emph{No Active Main Agent Monitoring}).

\noindent\textbf{B1. Wrong Start, Join, or Result Timing.}
This is the most frequent subcategory within execution governance. It occurs when sub-agent work starts, finishes, or is incorporated at the wrong time, causing required work or results to miss the final submission. A sub-agent may be terminated before finishing, finish too late for integration, or modify the artifact after verification. In the seqtk example (Figure~\ref{fig:illustrative-example-seqtk}), the main agent assumes a two-hour sub-agent timeout despite a 40-minute task budget. The task deadline therefore arrives before the planned join, terminating sub-agents with required work unfinished. Here, the delegation itself is reasonable, but its timing does not fit the available execution budget.

\noindent\textbf{B2. Failure Propagation.}
A sub-agent fails to complete its work, but the main agent neither retries, reassigns, nor takes over the unfinished work. In Claude Code's RepoZero canonicaljson test3 migration, a sub-agent assigned to search for argument-parsing counterexamples is interrupted when a transient request spike triggers rate limiting, but the main agent does not restart it or take over its verification task before timeout. The final artifact passes 147 of 148 official tests. The coordination problem is the unrecovered verification work.

\noindent\textbf{B3. No Active Main Agent Monitoring.}
The main agent fails to reassess progress and act when needed. It may keep waiting even when the task is already complete or fail to use available results. In Claude Code's RepoZero hopscotch-map test11 migration, the main agent has already built, checked, and placed the Rust port in the output directory. Near the deadline, however, it starts a ten-minute blocking wait for an unfinished sub-agent instead of finalizing the completed artifact. The execution is terminated before a final response, even though the artifact passes all 39 official tests. Unnecessary waiting prevents the main agent from closing an already completed task.

\subsubsection*{C. Global Context Management Problems}
It arises when agents working in separate contexts do not receive or preserve the context needed to connect their work. Specifically, a sub-agent may receive incomplete task requirements (\emph{Missing Task Requirements}), separately developed components may lack a shared interface contract (\emph{Missing Interface Contract}), information may be lost or become unusable during handoff (\emph{Overloaded Handoff}), the main agent may declare completion without checking all requirements (\emph{Unsupported Global Completion}), or a retry may repeat previous work without inheriting prior progress (\emph{Unchanged Retry}). Global context management accounts for 75 (9.3\%) of all identified pattern instances.

\noindent\textbf{C1. Missing Task Requirements.}
This subcategory occurs when a sub-agent does not receive all the requirements needed to complete its assigned work. In Kimi Code's NL2Repo icecream reconstruction, the task requires package metadata to be exported through \texttt{from icecream import *}, but the package sub-agent is not instructed to update the export list and returns an unchanged \texttt{\_\_init\_\_.py}. Although the artifact passes all tests, the delegation and recorded checks provide no evidence that this requirement was explicitly handled.

\noindent\textbf{C2. Missing Interface Contract.}
Different sub-agents develop related components without agreeing on how they should interact. In Kimi Code's NL2Repo python-fsutil reconstruction, implementation and test sub-agents receive separate file assignments without a complete shared API contract. The permissions sub-agent returns an integer, while the test-writing sub-agent expects strings such as \texttt{"755"}; other tests assume different function signatures. The first integrated test run consequently reports nine failures, requiring the main agent to reconcile these mismatches.

\noindent\textbf{C3. Overloaded Handoff.}
Information passed between agents is incomplete, distorted, or too large to be effectively used. In Claude Code's RepoZero base58 test7 migration, a Workflow aggregates up to 60 findings per probe into an 80K-character implementation brief. The receiving sub-agent nevertheless repeats environment, source, and traceback investigation instead of proceeding to implementation, and is interrupted before producing the required output file.

\noindent\textbf{C4. Unsupported Global Completion.}
The main agent declares the task complete without fully checking the integrated result. In Kimi Code's RepoZero bidict test19 migration, four sub-agents implement separate modules. After integration, the main agent checks stdout and exit codes but ignores stderr, even though it is aware of simplified exception output and an argument-parsing deviation. It nevertheless declares the task complete, leaving known requirements unverified.

\noindent\textbf{C5. Unchanged Retry.}
A replacement sub-agent repeats previous work without inheriting useful progress from earlier attempts. In Claude Code's RepoZero Clipper test5 migration, stalled design assignments to the sub-agent are repeatedly relaunched with the same brief. The replacements reread the source and repeat executable probes without receiving prior findings or a narrower next step. After four such attempts, the main agent reaches the deadline with only \texttt{Cargo.toml} and no required Rust entry file.

\subsubsection*{D. Shared State and Merge Problems}
These problems arise when concurrent agents read or modify shared files, artifacts, or environments without proper coordination, resembling data races in traditional concurrent programs. They form the largest main category, accounting for 267 of 804 (33.2\%) pattern instances.

\noindent\textbf{D1. Concurrent Writes.}
This subcategory accounts for 26.5\% of all pattern instances and 62.9\% of those identified for Codex. It happens when multiple agents modify the same or related files concurrently without coordination, creating write conflicts analogous to data races. In Codex's RepoZero Clipper test5 migration, the main agent writes and builds \texttt{test5.rs} and its supporting modules. A sub-agent later replaces \texttt{test5.rs} and \texttt{Cargo.toml} with a different module layout in the shared output directory. The main agent discovers the changes and must reconcile the mixed implementation before delivery. Although the final artifact passes the official test, the overwrite invalidates the main agent's prior view of the artifact and introduces unplanned integration work.

\noindent\textbf{D2. Shared Environment Contamination.}
Concurrent agents modify shared environment state without isolation, causing one agent to observe state unexpectedly changed by another. In Claude Code's clog\taskdash cli reconstruction, two sub-agents independently delete and recreate \texttt{/tmp/t1} for their own fixtures. After the Markdown sub-agent creates its fixture, the JSON sub-agent replaces the shared path, causing the Markdown sub-agent to unknowingly continue with the wrong fixture. In another instance, a sub-agent replaces the shared system Git executable, causing subsequent probes that invoke it within the same execution to use the modified executable.

\begin{takeawaybox}
\textbf{Finding 5:} Dynamic concurrency introduces distinct failure patterns beyond those in sequential agent execution. These patterns span task orchestration, execution governance, global context management, and shared state and merge, posing new challenges for agent reliability on long-horizon coding tasks.
\end{takeawaybox}

\FloatBarrier
\section{Benefit Analysis of Dynamic Concurrency (RQ4)}\label{sec:rq4}
We examine 131 pairs of concurrent and sequential execution trajectories produced by the same coding agent on the same task. Pairs are selected when concurrent execution achieves a higher test pass rate, fully solves a task that sequential execution does not, or reduces completion time by at least 60 seconds and 20\% without lowering test pass rates. By comparing both trajectories, we observe three recurring situations in which concurrency brings additional benefits for agents.

\noindent\textbf{Complementary Work.}
One common condition that concurrency can bring improvements is that sub-agents work on different and complementary parts of the entire task in parallel while the main agent successfully integrates their results. For example, in Kimi Code's NL2Repo justext reconstruction, eight sub-agents handle delivery files and checks while the main agent implements the core text-extraction algorithm. The main agent integrates all eight results before final verification. Though both concurrency and sequential modes pass all tests, concurrency reduces wall time from 734.116 to 467.997 seconds. Here, clearly separated responsibilities allow useful work to proceed in parallel and reduce overall completion time.

\noindent\textbf{Alternative Attempts and Selection.}
Dynamic concurrency can also be helpful when multiple agents develop different solutions to the same problem in parallel, allowing the main agent to select or combine the better results. In Codex's ProgramBench Halite reconstruction, the main agent and a sub-agent independently build replacements for the black-box game environment. The main agent combines the sub-agent's map behavior with its own stronger protocol handling, passing 292/391 tests, while the sequential mode fails to build and passes 0/391. Similarly, in NL2Repo unidecode, concurrent sub-agents provide implementations based on versions 1.3.8 and 1.4.0. After comparing them, the main agent adopts the 1.4.0 tables, passing more tests than sequential mode. In both cases, concurrency provides alternative solutions that improve the final implementation.

\noindent\textbf{Independent Validation.}
Dynamic concurrency is beneficial when sub-agents independently validate assumptions or implementations and provide evidence that helps the main agent correct mistakes. In Codex's LoopsBench security labs task, a crypto sub-agent identifies the P-256 curve-order requirement for ECDSA nonce reuse while the main agent implements the broader lab suite. The main agent uses this finding to generalize the recovery algorithm, and the concurrent artifact passes all checks, whereas the sequential mode fails. Similarly, in NL2Repo trimming, a nested sub-agent's upstream probe reveals the correct CSV header naming rules, thus the final artifact passes all tests but sequential mode cannot. In both cases, sub-agents created from concurrency provide corrective evidence early enough for the main agent to improve the final artifact.

\begin{takeawaybox}
\textbf{Finding 6:} Concurrency provides concrete benefits when parallel work is effectively used by the main agent: sub-agents can complete complementary work, explore alternative solutions, or independently validate and correct the main agent's implementation.
\end{takeawaybox}

\FloatBarrier
\section{Discussion}
\label{sec:discussion}

\subsection{Implications}

\noindent\textbf{Implications for researchers.} Our taxonomy of concurrency failure patterns reveals several open research problems for building reliable concurrent coding agents. First, \emph{task orchestration} requires better mechanisms for task decomposition, ownership assignment, and load balancing so that concurrency targets useful work without exhausting the execution budget~\cite{cataldo2008sociotechnical}. Second, \emph{execution governance} calls for runtime monitoring and intervention mechanisms that can detect stalled or failed sub-agents, recover unfinished work, and decide when to collect results or terminate execution. Third, \emph{global context management} motivates better ways to communicate task requirements, interface contracts, and reusable progress across isolated agent contexts. Finally, \emph{shared-state coordination} requires mechanisms for isolation, synchronization, and conflict detection when multiple agents concurrently modify shared files and environments~\cite{brun2011proactive}. Addressing these problems is important for turning dynamic concurrency from parallel agent execution into reliable collaborative software development.

\noindent\textbf{Implications for practitioners.}
Our findings provide practical guidance for practitioners on when to use dynamic concurrency: it should be enabled selectively rather than by default. In particular, dynamic concurrency is more promising for difficult, long-horizon tasks, whereas bounded or simpler tasks may see limited or even negative benefits. Practitioners should especially consider concurrency when a task can be clearly decomposed into independent components, supports multiple parallel attempts to the same problem, or allows independent sub-agents to validate the main implementation. At the same time, these potential benefits should be weighed against resource costs: dynamic concurrency substantially increases token usage and tool calls and often increases runtime, while performance gains remain uneven. Practitioners should therefore consider both task characteristics and available resource budgets when deciding whether and how aggressively to use dynamic concurrency~\cite{kim2026atropos,kim2025scalingagents,xiao2026agentdiet}.

\subsection{Threats to Validity}
\label{sec:threats}

\textit{External validity.}
We study three native concurrency mechanisms across five benchmarks covering diverse coding tasks and difficulty levels. Our findings are therefore scoped to the tested agents and benchmarks. To address this issue, we select the frontier coding agents with their native model across a wide range of benchmarks, leading to extensive evaluation costing over \$ 20K.
\textit{Internal validity.} Agent execution is stochastic, while the substantial cost of long-horizon evaluation limits our ability to perform large-scale repeated runs. Nevertheless, available evidence suggests that run-to-run randomness is unlikely to materially affect the overall trends observed in our study. In particular, Codex's sequential task and test pass rates on LoopsBench in our experiments are close to the originally reported results for Codex with GPT-5.4~\cite{li2026loopsbench}; moreover, the original LoopsBench study reports stable overall trends across repeated attempts~\cite{li2026loopsbench}.
Moreover, manual analysis may also introduce subjective judgments. For the failure analysis, we develop a codebook through a 50-trajectory pilot and have each case independently annotated by two of three annotators, with reasoning recorded for their label assignments. Annotators discuss disagreements, and a senior author adjudicates unresolved disputes. We define explicit coding criteria and inspect complete trajectories and artifacts before identifying a failure pattern.

\FloatBarrier
\section{Conclusion}
\label{sec:conclusion}

We study dynamic concurrency in three frontier coding agents through 2,124 concurrent and sequential executions across five benchmarks. Our results show that enabling concurrency increases token use while its effects on task success vary across tools and tasks. Trajectory analysis yields a taxonomy of concurrency failure patterns covering task orchestration, execution governance, global context management, and shared state and merge. We also identify three mechanisms in executions with observed advantages: complementary work, alternative attempts and selection, and independent validation. Across these cases, the main agent turns concurrent activity into useful progress by integrating components, selecting candidates, or applying corrective evidence. Our findings provide an empirical basis for designing and evaluating concurrency policies that prioritize coordinated, verified delivery under finite execution budgets.

\FloatBarrier
\section*{Data Availability}
\addcontentsline{toc}{section}{Data Availability}
\label{sec:data-availability}

The artifact is available at \url{https://github.com/schwerli/Concurrency-Failures-Trajectory-Artifact}. It contains the 2,124-cell manifest, official evaluator summaries, runner and parser versions, trajectory source hashes, normalized main agent and sub-agent event records, the qualitative codebook, review registers, and evidence for every reported case. Benchmark data and upstream repositories are referenced rather than redistributed where licenses require it. Scripts regenerate every numeric table and figure from frozen machine readable inputs.

\FloatBarrier
\bibliographystyle{ACM-Reference-Format}
\bibliography{references}

\end{document}